\documentclass{article}
\usepackage[a4paper, total={6.5in,9in}]{geometry}
\usepackage[utf8]{inputenc}
\usepackage{amsmath}
\usepackage{amssymb}
\usepackage{graphicx}
\usepackage{authblk}
\usepackage{url}
\usepackage{natbib}
\usepackage[ruled]{algorithm2e}
\usepackage{appendix}
\usepackage{rotating}
\usepackage{appendix}

\usepackage{longtable}

\usepackage[english]{babel}
\usepackage[autostyle]{csquotes}

\vspace{-40pt}
\title{Refining Understanding of Corporate Failure through a Topological Data Analysis Mapping of Altman's Z-Score Model}

\vspace{-30pt}
\author[1]{Wanling Qiu\thanks{\textbf{Corresponding Author}. Full Address: Accounting and Finance Subject Group, School of Management, University of Liverpool, 20 Chatham Street, Liverpool, L69 7ZH, United Kingdom. Tel: +44 (0)7955 109334 Email:wanling.qiu@liverpool.ac.uk}}
\affil[1]{School of Management, University of Liverpool, United Kingdom}
\vspace{-30pt}
\author[2]{Simon Rudkin \thanks{Full Address: Economics Department, School of Management, Swansea University, Bay Campus, Swansea, SA1 8EN, United Kingdom. Email:s.t.rudkin@swansea.ac.uk}}
\affil[2]{Economics Department, Swansea University, United Kingdom}
\vspace{-30pt}
\author[3]{Pawe{\l} D{\l}otko\thanks{ Full Address: Mathematics Department, College of Science, Swansea University, Bay Campus, Swansea, SA1 8EN, United Kingdom. Email:p.t.dlotko@swansea.ac.uk. }}
\affil[3]{Mathematics Department, Swansea University, United Kingdom}
\vspace{-40pt}
\vspace{-20pt}
\begin{document}
	\maketitle
	\begin{abstract}
		Corporate failure resonates widely leaving practitioners searching for understanding of default risk. Managers seek to steer away from trouble, credit providers to avoid risky loans and investors to mitigate losses. Applying Topological Data Analysis tools this paper explores whether failing firms from the United States organise neatly along the five predictors of default proposed by the Z-score models. Firms are represented as a point cloud in a five dimensional space, one axis for each predictor. Visualising that cloud using Ball Mapper reveals failing firms are not often neighbours. As new modelling approaches vie to better predict firm failure, often using black boxes to deliver potentially over-fitting models, a timely reminder is sounded on the importance of evidencing the identification process. Value is added to the understanding of where in the parameter space failure occurs, and how firms might act to move away from financial distress. Further, lenders may find opportunity amongst subsets of firms that are traditionally considered to be in danger of bankruptcy but actually sit in characteristic spaces where failure has not occurred.  
	\end{abstract}

	\maketitle
	
	Keywords: Credit Scoring; Topological Data Analysis; Data Visualization; Bankruptcy Prediction
	
	\section{Introduction}
	
	Credit default prediction models intuitively find direction from the financial fundamentals of the corporation and identify how such can be used to indicate likely future failures. In \cite{beaver1966} and \cite{beaver1968} individual financial ratios are tested for their ability to discriminate between firms that go bankrupt the following year and those who do not. \cite{altman1968financial} advances this to select the five ratios that best isolate potential failures and employs linear discriminate analysis to assign a coefficient to each. Subsequent developments charted for the 50th anniversary of the original \cite{altman1968financial} model in \cite{altman2017financial} have included extensions of the considered ratio set, considerations of non-linearity, removal of the normal distribution assumption implicit in the original multiple discriminate analysis, and the introduction of machine learning (ML) models. Such works break down into those that extend the information set and those which seek to extract more from the information already provided. Each extension has merit but with both comes the danger of over-fitting to particular values. More advanced techniques also bring questions of being a ``black box'' through which the link from input to output cannot be traced. It is then unsurprising that the original approaches continue to have resonance in the credit rating sector \citep{altman2017financial}.
	
	This paper returns to the fundamental models of \cite{altman1968financial} and it's predictions of default against the true cases of corporate failure. A Topological Data Analysis (TDA) Ball Mapper (BM) approach after \citep{dlotko2019ball} is used to produce an abstract two-dimensional representation of the financial ratio space to visualise where failures occur amongst the combinations of firm characteristics. Mapping the space in this way demonstrates how contemporary approaches in data science can break open the black box and illuminate how precisely the future of credit default prediction modelling should develop. Major advantages of the approach include a robustness to high levels of correlation between variables, noise within the dataset and critically respect the underlying data rather than imposing distributional assumptions.     
	
	Contributions of our work are thus threefold. Firstly, a demonstration of the application of a contemporary data science technique to financial analysis inspires a new understanding of the space upon which we conduct our evaluation of credit models. It is immediately seen that the areas of the characteristic space classified as likely to contain failures cover a lot of volume in which there are no failures. Secondly, it is demonstrated how non-linearity and interaction terms can all be accounted for in the evaluation of credit default risk; insights gained there from become invaluable for assessing firms. Finally a research agenda is signposted which can aid understanding of credit risk and open the ``black box'' of ML. TDA Ball Mapper thus offers much to the discussion of bankruptcy prediction.
	
	The remainder of the paper is organised as follows. Section \ref{sec:background} offers brief overview of the literature defining the problem of balancing fit against the risk of over-fitting. Section \ref{sec:data} details the data used for the empirical work, with Section \ref{sec:method} highlighting the TDA Ball Mapper approach and the theoretical expectations formed prior to the analysis. Through Section \ref{sec:results} and Section \ref{sec:discussion} a detailed review is undertaken as to whether the expected patterns are emerging. Finally, Section \ref{sec:conclude} concludes.
	
	\section{Literature Review}
	\label{sec:background}
	
	Corporate failure has obvious reprecussions for investors and those to whom the failed firm has liabilities. Consequently it is banks and financial institutions who are the greatest users of these models. For such users decisions on creditworthiness of potential borrowers must be traceable, clearly motivated by evidence and free from any allegations of black boxes. Whilst machine learning methods may generate more accurate predictions their data driven nature places too many unknowns in the process of getting from input information to output decision. Hence whilst there is a growing literature extholing the virtues of a machine learning approach this paper stands as a note of caution there against.
	
	\subsection{Development of Credit Default Models}
	
	\cite{altman1968financial} and \cite{edward1983corporate} models were constructed by multiple discriminate analysis (MDA) from a set of candidate factors. Each accounting ratio put forward as a potential explanatory variable is assessed for its ability to explain firm failure, with only those making a significant contribution being considered for the final model. The five factors that made the final \cite{altman1968financial} Z-score were chosen from a set of twenty-one. Later models from \cite{edward1983corporate} were designed to reflect non-listed firms who did not have a market value of equity to use in their evaluation. Such MDA approaches to identifying the drivers of firm failure dominated the literature to the turn of the century \citep{altman2017financial}. As well as MDA there was a growing thread after \cite{ohlson1980} application of logistic regression which employed probabilistic models. Extensions were made into other countries\footnote{See \cite{altman2017financial2} for a review.} but the fundamental ideas fell into one of these two categories. Critically, as demonstrated by the removal of asset turnover in the second of the \cite{edward1983corporate} frameworks because of differences between industries on this characteristic, the appraisal of the researcher was always maintained as a final check on the model produced.
	
	21st century work has been dominated by the growth of ML, with models seeking information from within the firm characteristic dataset through a variety of techniques. Many review studies chart this development, a good example being \cite{barboza2017}. In the early works two main families of model found favour. Support vector machines (SVMs) after \cite{cortes1995} have found credibility because they sit on the bridge between the MDA and ML, generating functions for credit scoring without the restrictions on functional form that regression necessitates \citep{altman2017financial}. However, the empirical work on European countries in \cite{altman2017financial} does not offer support to SVMs. \cite{petropoulos2016} contends that although SVMs have intuition on their side they remain ``black boxes''  that do not offer simple visualisations of the process through which they arrive at their predictions. Likewise the neural network ML models that also gained early prominence are openly critiqued for the inability of the practitioner to see the process through which the outcome was derived. Such a lack of transparency has consistently lead back to the MDA models. 
	
	Ensemble learning offers scope to combine the individual models to gain maximal prediction accuracy. In this way the algorithm is detecting what each approach is saying about the risk of the firm defaulting and then using that information through a weighted average to construct an overall expectation. Within the class of ensemble learning models lie bagging and boosting, which run the same algorithm multiple times and forming a linear combination from the outcomes.\cite{son2019} is a recent example of work on bankruptcy prediction using a boosting algorithm to extract more information from a neural networks model. Use of multiple algorithms is considered as ``hybrid'' ensemble learning. Early works in this area combined MDA and logistic models \citep{Li2012Forecasting}, while more recent work uses a fuller suite of ML models as well \citep{choi2018predicting}. \cite{debock2017} exposit how spline-rule ensembles can move learning beyond the linear combinations and, in so doing, address many of the non-linearities present within financial data. Unlike standard ensemble models these rule based ensemble learners can include different candidate algorithms at each node of the decision tree, different combination rules, and offer greater options for model interpretation.
	
	Financial practitioners are keen to know which variables are important, and what probability of default is associated with each level of the explanatory factors. \cite{debock2017} exposition of spline-rule ensemble learning shows how low cash ratios are associated with higher failure probabilities but increasing percentages of late payments being made by the company are unsurprisingly positively linked to failure. Some factors such as the solvency ratio and return on investment are non-linear. Such outputs, combined with the inherent increased accuracy that comes from combining measures, make ensemble learning models appeal to practice. However, the ``black box'' nature of the inputs to the ensemble set mean that the issue of traceability does not go away.
	
	\cite{ZiEnsemble} showcases another phase in the use of ensemble models whereby synthetic elements are introduced to understand outcomes. Such creation of artificial factors is common elsewhere in statistics and hence represents a natural extension to credit default risk. Such constructions help improve fit, and have intuition as the measurements that were not able to be captured in the real data, but are inherently consequences of the observed data points and hence potentially over-fit. 
	
	Financial default prediction models are thus becoming more accurate as data science offers ever improving means of extracting information from data. Whether through pattern recognition, probability of default function fitting, or simple classifications, the increased ability to use data improves fit relative to the \cite{altman1968financial} approach. However, it is imperative that understanding is driven from fundamentals and is clear to all users, to this end the ML model set have much to make up. It is to this requirement for transparency we speak.
	
	\subsection{Topological Data Analysis and Ball Mapper}
	
	In its most intuitive form a topological analysis seeks to create a map of a dataset, helping the analyst view what is going on in each part of the space. When constructing maps we intuitively turn to the longitude and latitude as point co-ordinates and then overlay information as contours, as colour or as points. This paper demonstrates a contemporary approach that generates a ``map'' of the financial ratios of firms. Rather than simply the two dimensions of the page the algorithm is constructing a two-dimensional visualisation of the multiple dimensional data that can then be placed onto the page. Other characteristics of maps, like colour and labelling can readily be overlaid. Because of the loss of dimensionality required to create the two dimensional map there is no longer any scale that allows measurement between points, but the full information remains to allow the computer so to do. From a practice perspective there is obvious value to such an approach.
	
	TDA has the immediate advantage of being constructed solely from the data collected. Once the variables to be collected are decided, measurements are taken and the data is plotted onto the space. By looking at the shape of the data TDA is constructing a map of what is there and not seeking to impose any relationships upon the data. Intuitively a linear model is assuming that there can be a straight line drawn through the data such that the outcome is a linear function of the input. Familiarity with adding regression lines to a scatter plot mean the effect of linear regression is understood. By using increasingly complex non-linear functions it is possible to fit the data better, reduce the residuals, but this comes at the cost of the model estimated being particular to the data in the sample. This is also known as over-fitting. Machine learning is helping to make that non-linear relationship more accurate to the data, but the over-fitting criticism still stands. Employing TDA enables understanding of the linearity of the relationship, and if there is no linearity allows the researcher to know more about the shape of the data. From a modelling perspective there are further obvious advantages created.
	
	Broadly TDA has been limited to the physical sciences where it is valued for its robustness to noise and ability to capture relationships between data points irrespective of the way in which they are differentially perturbed. For example in looking for genetic mutations it is important to be able to distinguish between small differentials between individuals and genuine changes in the gene that might signal a need for treatment\citep{nicolau2011}. Work to bring TDA into the finance field has typically focused on time series and considers the possibility of financial crashes \citep{gidea2018}. Therein it is the use of TDA to monitor for potential crashes in dynamical systems, such as production lines, that provides the inspiration. This paper represents one of the first applications of the cross section approach outside the natural sciences\footnote{At the time of writing the only known example is \cite{vejdemo2012topology}, which looks at voting behaviours in the United States of America House of Representatives.}. 
	
	From the perspective of the institution seeking to understand the riskiness of a particular business the value of the model based approaches is obvious. However, to really understand where the risk is high, drawing from the data is more intuitive. Taking the known characteristics of a firm and placing it within the picture can guide on the risk for such a firm. BM does not offer regression coefficients, but by looking at firms in the same space inference can be gained. Though analysts may have a feeling about which parts of space are risky, the BM algorithm may either confirm, OR sit at odds with, those initial thoughts. A process of learning what is really going on in the data is then the first step to getting the best impression of credit default risk.
	
	\section{Data}
	\label{sec:data}
	
	Data is constructed from Compustat and covers the period from 1961 through to 2015. Although there is more contemporary data available there are few recorded cases of failure since 2015 at the time of writing. This is due to the lag in cases entering the Compustat data. Formally a firm is regarded as failed if it either files for bankruptcy, or liquidates, in the financial year. For failed firms data from the most recent financial statements is provided alongside a deletion reason\footnote{Specifically this paper considers either bankruptcy (code 02) or Liquidation (code 03)}. 
	
	Explanatory variables are taken from the respective works of \cite{altman1968financial} and \cite{edward1983corporate}. Each is constructed from Compustat data using the formulae defined in Table \ref{tab:sumstats} and contains an allowance for size through a denominator of either total assets or, in the case of $X_4$, the total liabilities of the firm. These ratios capture the liquidity ($X_1$), profitability ($X_2$), productivity ($X_3$), leverage ($X_4$) and asset turnover ($X_5$). After winzorising at the 1\% level and removing any observations for which there is missing data, we are left with 110668 firm-years of which 3.7\% are failed firms.
	
	\begin{table}
		\begin{center}
			\caption{Variable Construction and Summary Statistics}
			\label{tab:sumstats}
			\begin{tabular}{l l l c c c c}
				\hline
				& Description & Compustat & Mean & s.d. & Min & Max \\ 
				\hline
				$X_1$ & Working Capital / Total Assets & $(act-lct)/at$ & 0.216 & 0.218 & -0.483 & 0.748\\
				$X_2$ & Retained Earnings / Total Assets & $re/at$& -0.089 & 0.900 & -7.699 & 0.694\\
				$X_3$ & EBIT / Total Assets & $(ni+xint+txt)/at$ & 0.041 & 0.170 & -1.014 & 0.310 \\
				$X_4$ & Market Value of Equity / Total Liabilities & $(csho \times prcc_f)/tl$& 2.943 & 4.216 & 0.093 & 31.92\\
				$X_5$ & Sales / Total Assets & $sale/at$ & 1.119 & 0.704 & 0.001 & 3.542\\
				& Firm Failure & $delrsn = 1,2$ & 0.037 & 0.188 & 0 & 1\\
				\hline
			\end{tabular}
		\end{center}
		\raggedright
		\footnotesize{Notes: All data is sourced from Compustat. Description provides the formulae from \cite{altman1968financial} or \cite{edward1983corporate} for the construction of the $X$ variables. The column Compustat details the variable names used in the construction of the explanatory factors ($X_1$ to $X_5$). Compustat variable names are as follows $act$ - current assets, $lct$ - current liabilities, $at$ - total assets, $re$ - retained earnings, $ni$ - net income, $xint$ - interest payments made, $txt$ - taxation on earnings paid, $chso$ - current shares outstanding, $prcc_f$ - price of the share at the financial year end, $sale$ - total sales of the firm and $delrsn$ is the reason for deletion from the Compustat database.  Firm failure is a dummy for deletion from the Compustat dataset in the subsequent year owing to either bankruptcy or liquidation. Sample from 1961 to 2015, $n=110668$}
	\end{table}
	
		\begin{table}
		\begin{center}
			\caption{Full Sample Correlations}
			\label{tab:cor}
			\begin{tabular}{l c c c c c c}
				\hline
				& $X_1$ & $X_2$ & $X_3$ & $X_4$ & $X_5$ & Fail \\ 
				\hline
				$X_1$ & 1 &&&&&\\
				$X_2$ & 0.108 & 1 &&&&\\
				$X_3$ & 0.128 & 0.511 &1 &&& \\
				$X_4$ & 0.279 & -0.021 & 0.136 & 1 &&\\
				$X_5$ & 0.355 & 0.066 & 0.102 & -0.055 & 1 &\\
				Fail & 0.040 & -0.010 & -0.046 & -0.043 & 0.046 & 1\\
				\hline
			\end{tabular}
		\end{center}
		\raggedright
		\footnotesize{Notes: All data is sourced from Compustat. Financial ratios are $X_1$ (liquidity), $X_2$ (profitability), $X_3$ (productivity), $X_4$ (leverage) and $X_5$ (asset turnover). Fail is a dummy for deletion from the Compustat dataset in the subsequent year owing to either bankruptcy or liquidation. Sample from 1961 to 2015, $n=110668$}
	\end{table}
	
	Table \ref{tab:cor} provides full sample correlations between the five explanatory ratios of \cite{altman1968financial} and the firm failure dummy. Amongst these there are high correlations between profitability, $X_2$, and productivity, $X_3$, as well as between liquidity , $X_1$, and asset turnover, $X_5$. None of these values touch the 0.7 in absolute value that would be seen as a sign of multicollinearity, but the correlation is high and so regression analyses should note that in their exposition.
	 
	It is recognised that effects may vary from year to year and so Table \ref{tab:yearsum} summarises the five ratios for each year considered in this paper. For brevity the number of years considered is just 6; being in 10 year intervals to the most recent data and 2008 to capture what was happening at the start of the global financial crisis. Failure proportions were much higher during the early years, whilst most recently the failure proportion has been very low. Even at the height of the financial crisis the percentage of firms that failed was just 1\%. 
	
	\begin{table}
		\begin{center}
			
				\caption{Annual Summary Statistics}
				\label{tab:yearsum}
				\begin{tabular}{l c c c c c c }
					\hline
					Year & \multicolumn{5}{l}{Financial Ratios} & Failure \\
				    & $X_1$ & $X_2$ & $X_3$ & $X_4$ & $X_5$ &  (\%) \\
					\hline
					
					1975&0.273&0.259&0.111&1.229&1.169&6.91\% \\
						&(0.196)&(0.250)&(0.091)&(1.792)&(1.721)&\\
					1985&0.234&0.073&0.045&2.492&2.519&6.26\%\\
					&(0.22)&(0.544)&(0.169)&(3.497)&(4.105)&\\
					1995&0.214&-0.101&0.033&3.682&3.735&2.90\%\\
					&(0.225)&(0.779)&(0.174)&(4.951)&(5.541)&\\
					2005&0.196&-0.361&0.021&4.081&4.076&2.01\%\\
					&(0.224)&(1.222)&(0.180)&(4.992)&(5.118)&\\
					2008&0.188&-0.327&-0.013&2.446&2.278&1.01\%\\
					&(0.225)&(1.181)&(0.216)&(3.796)&(3.738)&\\
					2015&0.17&-0.370&-0.013&3.072&3.163&0.03\%\\
					&(0.214)&(1.228)&(0.191)&(4.225)&(5.485)&\\
					\hline
				\end{tabular}
			
		\end{center}
	\raggedright
	\footnotesize{Notes:Financial ratios following \cite{altman1968financial} are $X_1$ (liquidity), $X_2$ (profitability), $X_3$ (productivity), $X_4$ (leverage) and $X_5$ (asset turnover). Failure classified as de-listing from Compustat owing to either bankruptcy or liquidation. Data from Compustat.}
	\end{table}

	\section{Methodology}
	\label{sec:method}
	
	Analysis of the shape of the data begins with the construction of the point cloud. For credit default modelling using \cite{altman1968financial} this is achieved by plotting each firm as a point in five dimensions. Each coordinate being one of the $X_j$'s used in the formation of the Z-score. In this paper different clouds are formed for each of the years studied. Firms which are proximate in the five dimensional space must have similar values for all of the considered financial ratios. A theoretical introduction to the method follows, with consideration then given to the representation that might be expected to emerge.

\subsection{TDA Ball Mapper}
Representation of the multi-dimensional point cloud is achieved using the BM algorithm of \cite{dlotko2019ball} as implemented in the R package \textit{TDABallMapper} \citep{dlotko2019R}. There are a number of advantages of the \cite{dlotko2019ball} approach over the original mapper algorithm developed by \cite{singh2007topological} and implemented in the \textit{TDAmapper} package of R \citep{tdamapper}. These advantages may be briefly summarised in the consistency of representation of the point cloud created by the BM algorithm, and its use of a single parameter rather than the more complex inputs required to the original approach.

Formally, the BM algorithm of \cite{dlotko2019ball} starts with the firm characteristic point cloud  $X$ and a constant $\epsilon > 0$. It select a subset $C \subset X$ having the property that the set of balls $B(C) = \bigcup_{x \in C} B(x,\epsilon)$ contains the whole set $X$. Such a subset $C$ is referred to as an $\epsilon$ net. Algorithm 1 of the \cite{dlotko2019ball} paper identifies neatly how the $\epsilon$-net $C$ is formed.

\begin{algorithm}[H]
	\SetAlgoLined
	\KwIn{Point cloud $X$, $\epsilon>0$}
	$C = \emptyset$\;
	Mark all points of $X$ as \emph{uncovered}\;
	\While{There exist uncovered $p \in X$ }{
		$C = C \cup p$;\;
		Mark every point $x \in B(p,\epsilon)$ as \emph{covered}\;
	}
	\KwOut{C}
	\caption{Greedy $\epsilon$-net\citep{dlotko2019ball}}
\end{algorithm}

In this construction the ball radius $\epsilon$ is the only exogenous input. Choosing $\epsilon$ recognises the competing forces of maintaining detail and producing a representation upon meaningful inference can be made. The sequential process of Algorithm 1 can produce slightly different results based on the random selection of the next uncovered point $p$, but because all possible $\epsilon$ nets are close to each other, the impact of this randomness to the overall output is marginal.
Owing to the way that the balls are formed the maximal distance of points from the ball's centre is bounded by $\epsilon$. This may be the entire $\epsilon$ on one axis and zero distance on others. The total quantity of the distance can also be shared out across all the axes. To think about this consider the unit circle centered at the point $(0,0)$ being drawn on a two dimensional plane. In this case, every point $(x,y)$ that satisfy $\sqrt{x^2+y^2}\leq 1$ will be in that unit circle.
In general it is not possible to decide the optimal value of the parameter $\epsilon$ and so it is left to the researcher to determine how big to set the radius. 

\begin{algorithm}
	\SetAlgoLined
	\KwIn{$C, X, \epsilon$}
	$V$ = abstract vertices, one per each element of $C$\; 
	$E = \emptyset$, \;
	\For{$p_1,p_2 \in C$ such that there exist $x \in X \cap B(p_1,\epsilon) \cap B(p_2,\epsilon)$}
	{
		$E = E \cup \{p_1,p_2\}$ \; 
	}
	
	\KwResult{BM Graph, G=(V,E)}
	\caption{Construction of a BM graph \citep{dlotko2019ball}}
\end{algorithm}

Conversion of the output from Algorithm 1 into a TDA Ball Mapper graph requires a further stage of graph construction. Algorithm 2\footnote{This is Algorithm 3 in \cite{dlotko2019ball}.} provides such a step  when an abstract graph to summarize the shape of $X$ is constructed. As defined by the algorithm an edge is drawn between the centroids of every two balls which have data points in their intersection; such lines helping identify where in the cloud each ball sits relative to the others.  Because of the way that the graph is constructed it would be expected that more vertices would appear in the BM than when using conventional mapper. Consequently, there may be additional information which is visualised in the BM graph. 

An important decision in the construction of graphs is whether the variables should be normalised. In this paper we do normalise all axes onto the range $[0,1]$ to recognise the variability in ranges identified in Tables \ref{tab:sumstats} and \ref{tab:yearsum}. In other applications normalisation might not be appropriate.
	
	\subsection{Interpreting TDA Ball Mapper Graphs}
	
	\begin{figure}
		\begin{center}
			\caption{Interpreting Ball Mapper graphs}
			\label{fig:bmeg}
			\includegraphics[width=10cm]{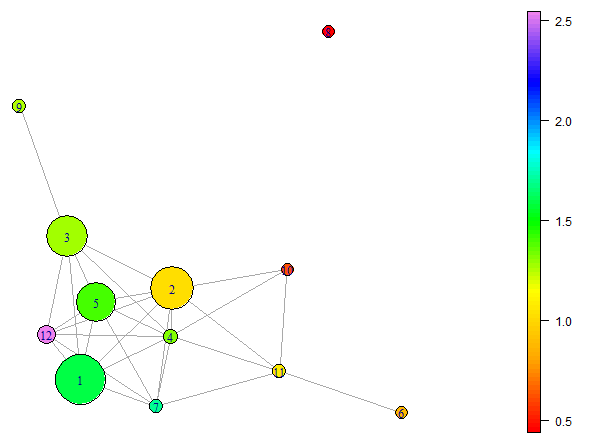}
		\end{center}
	\raggedright
	\footnotesize{Notes: Example TDA Ball Mapper plot created using \textit{BallMapper} \citep{dlotko2019R}. Axes following \cite{altman1968financial} are $X_1$ (liquidity), $X_2$ (profitability), $X_3$ (productivity), $X_4$ (leverage) and $X_5$ (asset turnover). Data is from Compustat and represents the value of these variables in 1975. Colouration is the Z-score as calculated using equation \eqref{eq:orig}. All axes are normalised to 0,1. $\epsilon=0.5$. }
	\end{figure}
	
	BM graphs have several key features that aid understanding the data they plot. Although necessarily abstract, the BM graph does maintain topological faith to the underlying dimensions. As an illustration of the properties a plot using 1975 data is provided in Figure \ref{fig:bmeg}.
	
	Firstly, the colouration of the graph allows analysis of the distribution of an outcome of interest across the space. This may simply be the average value for all the data points contained within the ball, as is done in this paper, but it is also possible to use counts, standard deviations, minima, maxima, etc. The choice of function is methodologically left to the user to define. Because in most instances it is the average outcome that is considered most representative of a ball, it is this which is the default function in the \textit{BallMapper} package. A scale to the right of the plot shows the values, here Z-scores from the \cite{altman1968financial} model. In this way it can be seen that the lowest scores sit to the right of the plot, with the only ball averaging over 2 being ball 12. 
	
	Secondly, the size of the balls gives indication of the number of data points located within that part of the plot. Bigger balls mean more points and a denser data concentration within that $\epsilon$ radius of the central point of the ball. In Figure \ref{fig:bmeg} it is ball 1 that has the most points contained within it, closely followed by 2, 3 and 5. There are a number of less populated balls spreading out to the right of the figure. Some like 4 and 12 are very close to these larger balls.
	
	Connectivity between balls represents the third useful feature of a BM graph. Figure \ref{fig:bmeg} contains a large number of edges emanating from most of the balls. In this way we would conclude that there is a lot of overlap and hence the whole graph is covering a cloud of similar data. There are however smaller arms sticking off to balls 6 and 9 that would represent parts of the cloud that extend out from the main collection of points. Finally where a ball is not connected we are identifying potential outliers. Ball 8 to  the top right of Figure \ref{fig:bmeg} is such a point. Note that the location of any disconnected balls does not show where they sit relative to the main connected components since the plot is abstract.
	
	Fourthly the BM plots demonstrate correlation between the axis variables. For example in the two dimensional case a set of correlated points occur within a narrow band, this extends into multiple dimensions such that the BM graph will itself be closer to being a long thin shape. Note that the BM graph is abstract so it is unlikely that the line drawn would be truly straight, but rather it would bend round to fit within the plot. Where variables are less correlated the cover will need to spread out and the more net appearance seen at the lower left of Figure \ref{fig:bmeg}. 
	
	Although not reported immediately by the plot, the number of balls will give the analyst a feeling for the level of detail contained. Smaller ball radius parameters, lower $\epsilon$, will lead to more balls being needed to cover the set of data points. Precise determination of $\epsilon$ for any given application is a matter for the analyst to determine, but we might conclude that the choice made in Figure \ref{fig:bmeg} is too high as there is not much detail being gained at the centre of the plot. In what follows a smaller $\epsilon$ is used.
	
	BM thus has a number of useful features that can help interpret the link between firm characteristics and firm failure. As with all methodologies the final choice of inputs will be the defining factor for the value of the analysis performed.

\section{\cite{altman1968financial} Z-score Model}
\label{sec:results}

\cite{altman1968financial} proposed the Z-score model for predicting firm failure as:
\begin{align}
Z&=0.012X_1+0.014X_2+0.033X_3+0.006X_4+0.999X_5 \label{eq:orig}
\end{align}
concluding that a Z-score of larger than 2.99 would place the firm in the safe zone and unlikely to suffer distress. A Z-score between 1.8 and 2.99 places the firm in a ``grey'' zone where failure cannot be ruled out. Should the Z-score be below 1.8 then \cite{altman1968financial} assigns the firm to a ``distress'' zone. 

\begin{figure}
	\begin{center}
		\caption{Altman Z Scores and Firm Failures: Original Model 2015 ($\epsilon=0.4$)}
		\label{fig:on2015}
		\begin{tabular}{p{2.5cm} p{2.5cm} p{2.5cm} p{2.5cm} p{2.5cm} p{2.5cm}}
			\multicolumn{3}{c}{\includegraphics[width=7cm]{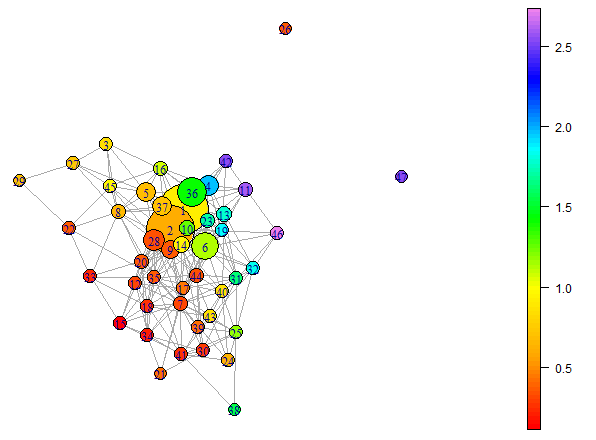}}
			&
			\multicolumn{3}{c}{\includegraphics[width=7cm]{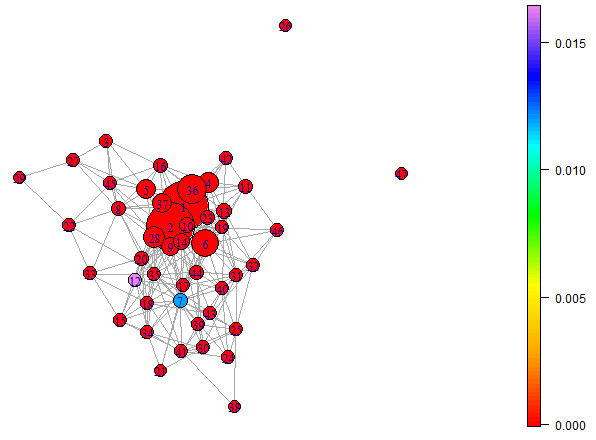}}
			\\
			\multicolumn{3}{c}{(a) Z-scores} & \multicolumn{3}{c}{(b) Failure proportion} \\
			\multicolumn{2}{c}{\includegraphics[width=5cm]{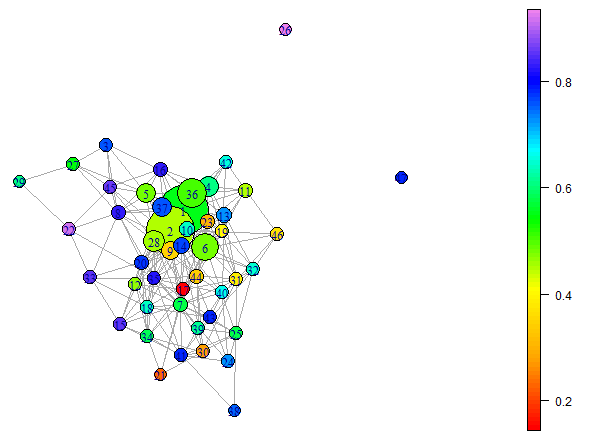}}&
			\multicolumn{2}{c}{\includegraphics[width=5cm]{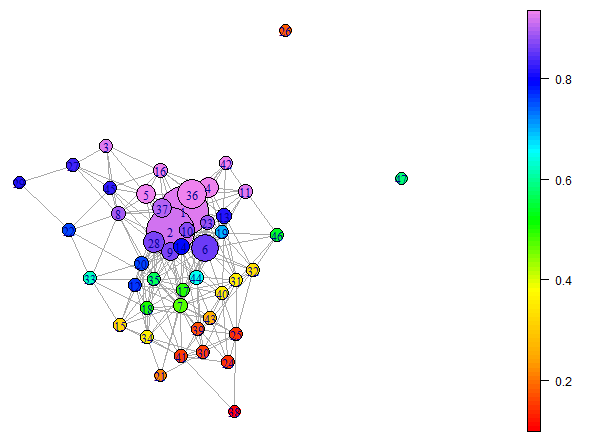}}&
			\multicolumn{2}{c}{\includegraphics[width=5cm]{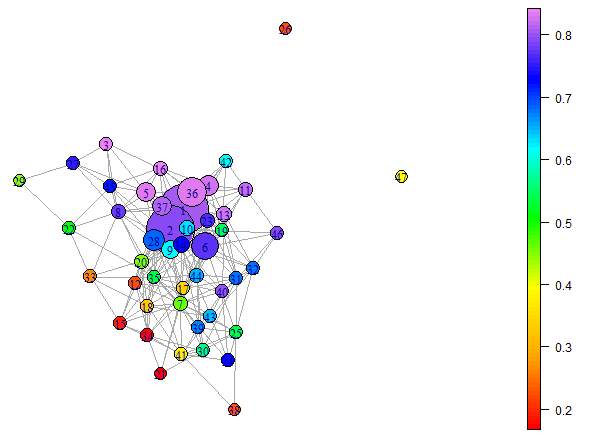}}\\
			\multicolumn{2}{c}{(c) $X_1$} & \multicolumn{2}{c}{(c) $X_2$} & \multicolumn{2}{c}{(c) $X_3$}\\
			\hspace{2.5cm}& \multicolumn{2}{c}{\includegraphics[width=5cm]{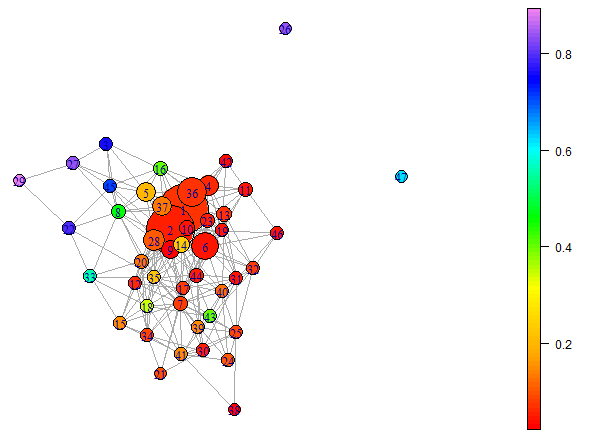}}&
			\multicolumn{2}{c}{\includegraphics[width=5cm]{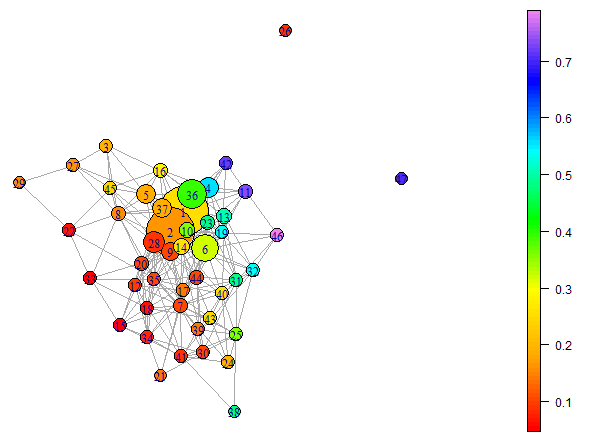}}&\\
			& \multicolumn{2}{c}{(c) $X_4$} & \multicolumn{2}{c}{(c) $X_5$} &\\
			
		\end{tabular}
	\end{center}
\raggedright
\footnotesize{Notes: TDA Ball Mapper \citep{dlotko2019ball} plots of the five dimensions of the original \cite{altman1968financial} model generated using \cite{dlotko2019R}. Axes are $X_1$ (liquidity), $X_2$ (profitability), $X_3$ (productivity), $X_4$ (leverage) and $X_5$ (asset turnover). All axis variables are normalised to the range 0 to 1 for consistency. Panel (a) is coloured according to the z-score calculated by equation \eqref{eq:orig}. Panel (b) is shaded according to the proportion of observations within the ball that did fail. Panels (c) to (g) are coloured based upon the variables used in the construction of the Z-score. Here we the abstract nature of the plots. Diversity in colour stems from the normalisation process as evidenced in comparison with the actual value plots. Figures are available in colour in the on-line version of the paper.}
\end{figure}

As an illustration of the benefit of TDA consider the space defined by the original Altman model. BM is used to construct an abstract representation of the firm characteristics space, here in the five dimensions set out as $X_1$ to $X_5$ in Table \ref{tab:sumstats}. In this way the segments of the point cloud with low values for the Z-score will be clearly visible. It is then asked whether the firms that failed were indeed in this part of the space. If the model is effective then it would be expected that the proportion of firms within a given ball that fail would be highest in the balls identified as having low Z-scores. Further it should follow that the proportion of failure in balls with high Z-scores should be 0. Figure \ref{fig:on2015} examines this for 2015. To construct a BM plot a ball radius must be selected. For this purpose $\epsilon=0.4$ is used, meaning that if all other variables are the same the largest distance between two points in a ball on the single differential axis will be 0.8 in the normalised space. This may seem large but since there are five axes that means the average distance from the central point is just $0.4/5=0.08$. Moving to lower numbers produces a very large set of balls and makes inference more challenging\footnote{Results from other filtrations are available on request from the authors.}.

Figure \ref{fig:on2015} is divided into three key parts. Firstly the Z-scores predicted using equation \eqref{eq:orig} are plotted in panel (a). Lower values, associated with predictions of failure, are located to the bottom left of the plot and are denoted by reds and oranges in the shading. Higher values are found to the right and towards the top represented by the blues and purples. In \cite{altman1968financial} a Z-score below 1.80 is considered as placing a firm in the ``distress'' zone. In the plot the ``distress'' zone will also include the big balls at the centre right. Panel (a) also reveals that no ball has an average Z-score above 2.99 meaning that no ball is considered entirely within the ``safe'' zone. Panel (b) is coloured according to the proportion of firms within a given ball that suffer failure in the following year. The highest proportion is 1.5\% and occurs to the lower left of the big mass. A comparison with panel (a) shows that these were indeed low Z-score balls. In this way the plots are suggestive that the original \cite{altman1968financial} model does identify potential failures in 2015. However, the model states that all those with Z-scores below 1.8 should be considered as being in the ``distress'' zone and it is clear that it is only a subset of these firms that did go on to fail. To diagnose the source of this colouration according to the five axis variables may be undertaken.

Correspondence between the failure proportions and the values of $X_1$ is much closer, both of the balls with higher proportions of failure correspond to higher values of $X_1$. In \eqref{eq:orig} the coefficient on $X_1$ is small and so it has a much smaller effect on the Z score. There is some evidence in these plots that a higher coefficient would be beneficial to represent the 2015 data. $X_2$, shown in panel (d), is higher towards the top. There is some correspondence with the overall Z-score, but where the latter is low on the top left $X_2$ is not. Explaining the low Z-Score in the top left is best done by looking at $X_3$. Indeed panel (e) confirms this. Panel (g), $X_5$ also has a strong correlation with the Z-Score with higher values in the top right. Panel (f) shows a much greater diversity of spread for $X_4$, only in the top left of the plot is any consistency observed. Failure in 2015 appears to be most associated with $X_1$, but again there are a lot of high values of $X_1$ where failure is not seen. 

An immediate observation from the 2015 data is that there are many stories behind the data which the linear discriminate models are not bringing out. Indeed the use of the variables without interactions would still not be able to make the discriminations suggested by Figure \ref{fig:on2015}.

\begin{figure}
	\begin{center}
		\caption{Altman Z Scores and Firm Failures: Original Model}
		\label{fig:on1985}
		\begin{tabular}{c c c}
			\includegraphics[width=5cm]{a5plot2015eps140normorigz1a.png}&
			\includegraphics[width=5cm]{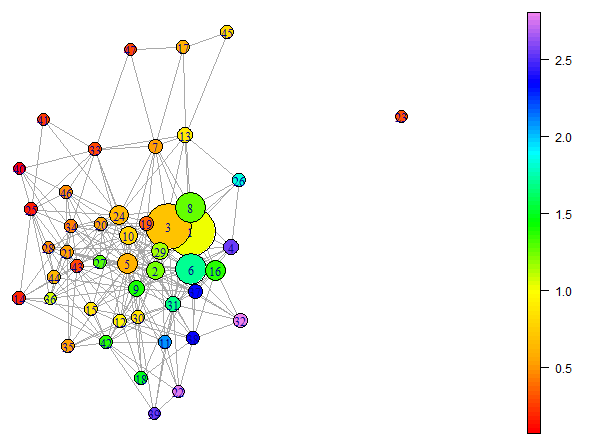}&
			\includegraphics[width=5cm]{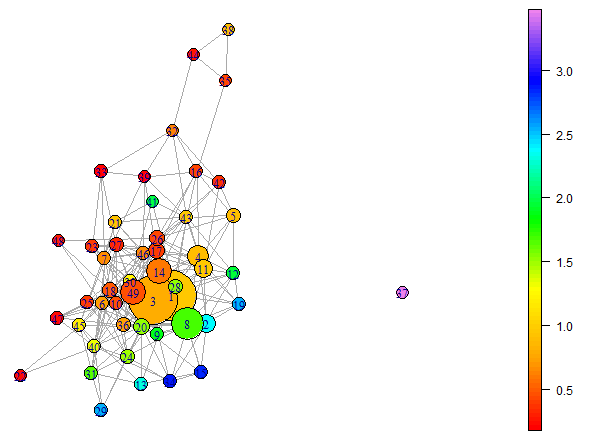}
			\\
			(a) 2008 Z-score & (b) 2005 Z-score & (c) 1995 Z-Score\\
			\includegraphics[width=5cm]{a5plot2015eps104normorigz1afail.png}&
			\includegraphics[width=5cm]{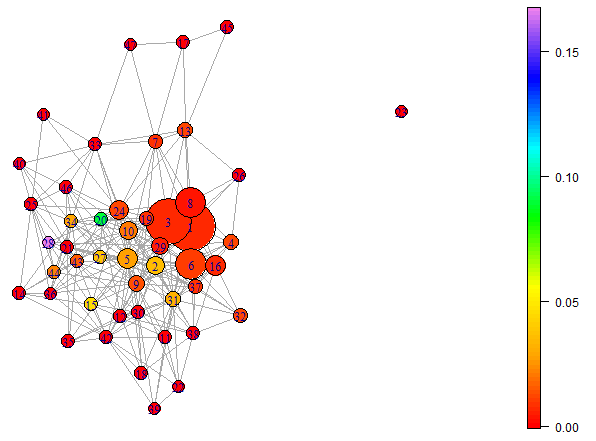}&
			\includegraphics[width=5cm]{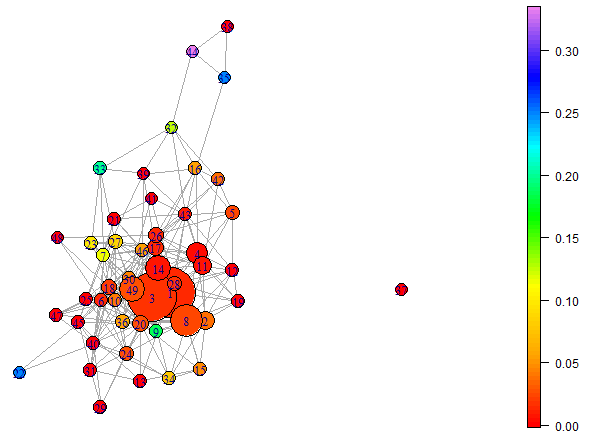}
			\\
			
			(d) 2008 Failure proportion & (e) 2005 Failure proportion & (f) 1995 Failure proportion\\
			
			\includegraphics[width=5cm]{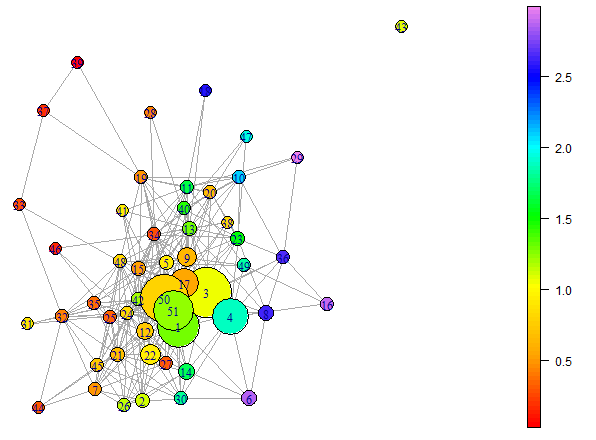}&
			\includegraphics[width=5cm]{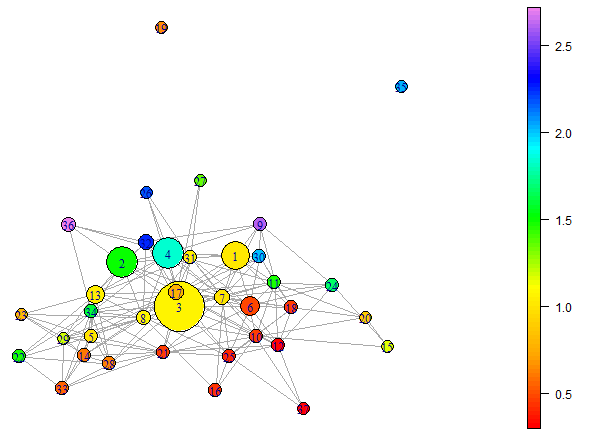}&
			\includegraphics[width=5cm]{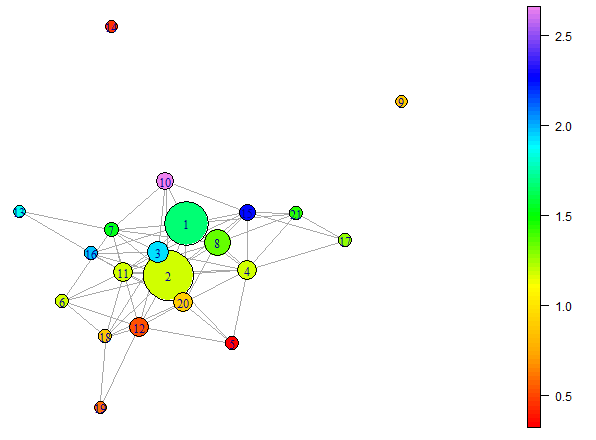}\\
			(g) 1995 Z-score & (h) 1985 Z-score &  (i) 1975 Z-score\\
			\includegraphics[width=5cm]{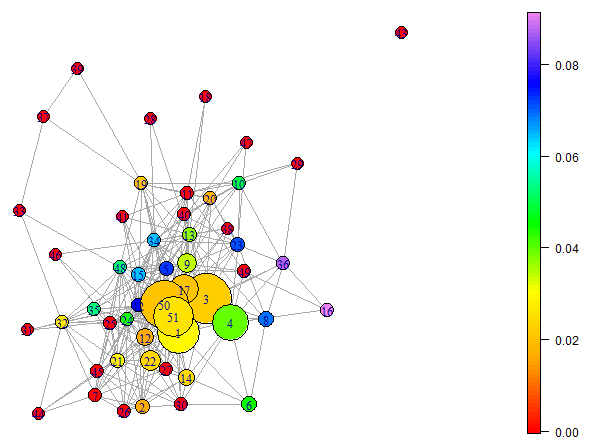}&
			\includegraphics[width=5cm]{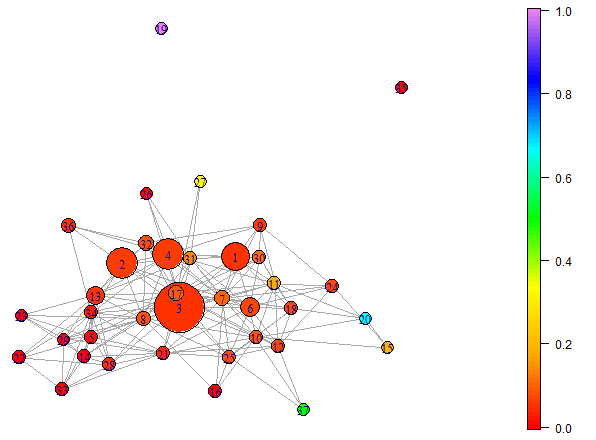}&
			\includegraphics[width=5cm]{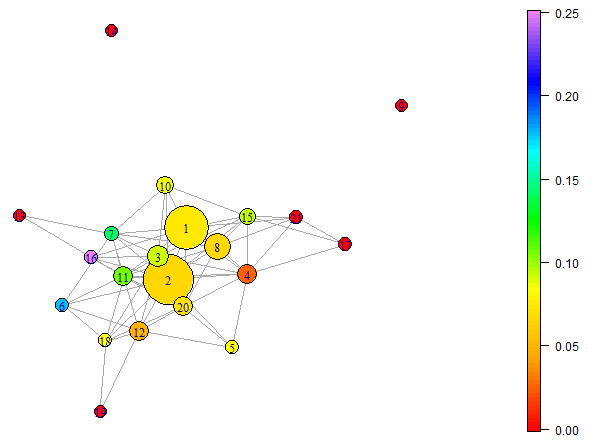}\\
			(j) 1995 Failure proportion & (k) 1985 Failure proportion & (l) 1975 Failure proportion\\
			
		\end{tabular}
	\end{center}
	\raggedright
	\footnotesize{Notes: TDA Ball Mapper plots generated using \cite{dlotko2019ball} for the original \cite{altman1968financial} model. Axes are $X_1$ (liquidity), $X_2$ (profitability), $X_3$ (productivity), $X_4$ (leverage) and $X_5$ (asset turnover). Panels (a) to (c) and (g) and (i) are coloured according to the z-score calculated by equation \eqref{eq:orig}. Panels (d) to (d) and (j) to (l) are shaded according to the proportion of observations within the ball that did fail. Figures are available in colour in the on-line version of the paper.}
\end{figure}

Appraising the fit of the original model in older years, Figure \ref{fig:on1985} has the Z-score and failure proportion plots for 1975, 1985, 1995, 2005 and the height of the global financial crisis in 2008. All plots have a similar lattice format to the 2015 case. However, a narrower, longer, shape in 2005 suggests that the data was more correlated that year. It has already been noted that a strength of the TDA Ball Mapper approach is that it can continue to be applied in cases like this.    

Panels (a) and (d) of Figure \ref{fig:on1985} show the Z-scores and firm failure proportions for 2015 and are included for reference. Interest begins with panels (b) and (e) which show these two outcomes for 2008. This was the start of the global financial crisis so theoretically may have the most surprising exits from the Compustat database. Compared to the 2015 plot only a few firms are obtaining the highest Z-scores. The region coloured green, yellow, orange and red, covers the majority of the space. The largest balls in the plot are also in the distress region according to the Altman Z-score. Failures are indeed seen across the space, but the larger proportions are concentrated in the lower centre of the graph. Again this informs that an interaction between the variables will be better to identify where exactly exiting the Compustat listing will occur. 

For 2005 panels (c) and (f) reveal a similar story of the ``distress'' zone covering a much larger proportion of the space than the other levels. There are failures in the lower part of the plot that correspond with the high Z-scores and low Z-scores. The most intense of the failure proportions appear in the top of the plot, far from the high Z-score end. Panels (g) and (j), plotting 1995, show that there is a smaller coverage of balls with low Z-scores. Failure proportions in panel (j) here correspond more with the high Z-scores to the right of the plot; the \cite{altman1968financial} model does not perform well for 1995. By contrast panels (h) and (k) for 1985 have the failures primarily concentrated in the bottom right, an area with very low Z-scores. There are also low failure proportions within the biggest balls, here again the average Z-score is well below the 1.8 cut off for the ``distress'' zone. Going back through time the same filtration produces fewer balls, the 1975 plots in panels (i) and (l), are particularly simplified relative to the others. Here the failures sit to the left of the plot in an area where there were some very high Z-scores noted. 

Overall the TDA Ball Mapper plots have usefully shown that the firms that failed have characteristics in, or around the boundary of, the ``distress'' zone. This will explain the high accuracy of prediction from the \cite{altman1968financial} model. However, there are also many cases where the failed firms sat in areas of the plot where Z-scores were high and financial distress was not expected. Two important messages thus emerge. Firstly there is a need to split the ``distress'' zone using the interactions between variables. Secondly, non-linearity between the financial ratios and outcomes apply across the space. 

\section{Implications and Discussion}
\label{sec:discussion}

This paper seeks to evaluate the effectiveness of credit modelling in predicting firm failure, the lens applied making it possible to identify segments of the parameter space in which failures occur and, through the functionality of the \textit{BallMapper} package, to identify the intensity of the failure rate in any given part of the space. In illustrating the power of the approach a simple dataset just using the five financial ratios of \cite{altman1968financial} was used. Therein lie a number of stories that create non-linearity in the link between the characteristics and failure; failures are not distributed evenly across that part of the space where the \cite{altman1968financial} model suggests financial distress. An overwhelming message from the analysis is that there is non-linearity and a clear importance to looking at the interactions between variables.

\cite{edward1983corporate} suggested that the Z-score can be a useful means of informing managers about the ways in which they may turn their business around. The BM graph informs where a company sits in the space and hence where the nearest balls that are not associated with failure are. Using the information in the BM graph it is straightforward to look at which variables would have placed the firm in a safer part of the space and hence to understand the most effective route away from potential failure. In most instances this will mean moving along one of the edges of the graph since already there are overlaps that imply proximity of the two ball centres.

Causality is not discussed within the context of the BM graph. What is seen is the actual occurrence of firm failure in the year after the Compustat financial ratios are observed. That there are no failures in a ball in a given year does not mean that it would have been impossible for a firm in that ball to fail. Likewise it is certainly not the case that any firm that finds itself in a ball where there have been failures will itself fail. In the examples provided here there is one case where a ball has 100\% failure, but typically the rates are less than 25\%. It can then be considered whether to build a causal model using the inference from the BM graph to guide variable inclusion.

Motivation for the use of BM came from the non-linearity criticisms of the MDA papers after \cite{altman1968financial}. Linearity is also a feature of many of the probabilistic models that follow \cite{ohlson1980} also. By plotting the space using BM there is clear evidence that the failure outcome is not distributed evenly across the space in the way that a linear function, like the Z-score of \cite{altman1968financial} is. Including all of the interaction terms would identify the differences in the space and may be the next step for the construction of a better model. However, such inclusion of the full set of interactions becomes much harder as the original number of variables increases. To have all the interactions for the five axes used here necessitates a further 24 variables on top of the original 5. As the number of explanatory factors increases so there is a rapid increase in the number of parameters to be estimated. Depending then on the complexity of the modelling such large volumes of unknowns may be problematic for simultaneous explanation.

Simple models additionally struggle when the data that is being used is highly correlated. Post normalisation the BM graphs do not have the narrow shape associated with strong correlation but as seen in the summary statistics the data does show strong correlation between some of the variables. Were the set of explanatory ratios to be expanded further then potential multicollinearity is a bigger issue and the benefits of using BM would, as with the way that the number of interactions increases with potential factors to include, become more pronounced.

Contemporary growth in credit modelling has been driven by ML modelling, seeking to bring the information from data through the search of specific patterns, classifications or simply the ability to fit non-linear functions. Motivating these advances are improved fit that comes from being able to use models that are much freer in their functional form. Recent advances in ensemble learning and the use of synthetic factors have served to improve model fit even further. However, all techniques have been subjected to criticism for their ``black-box'' nature because the precise details of the model fitting are not available in the way that the MDA and logistic models are. By looking at the stark difference between the predicted Z-score and the observed failure the BM graphs have shown that to get a better fit it is necessary for these ML approaches to do a large amount of deviation from the linear models. Hence the extent to which the unobservable functional form is relevant is more pronounced. 

From a practitioner perspective the evidence presented here can help with either the decision to lend, or, in the consideration of how best to escape from financial trouble. On the former a potential lender could identify where in the space the applicant for credit is, and therefore make a decision on how likely that candidate is to slip into distress. Presuming lenders gain profit from all successfully repaid loans this would open up huge parts of the ``distress'' region of \cite{altman1968financial} that would actually be solid opportunities. This is the motivation of the softspace clustering algorithms discussed in \cite{liu2019}; in that literature strand clustering algorithms are used to segment the space and different models are constructed for each cluster. Producing models for each ball would be an option but brings back the danger of over-fitting past data. Developing this strand, whilst maintaining the necessary generalizable results, will be an important next step. Having a location on the map for a candidate firm permits manual analysis of risk given positioning in the data cloud and remains an important, model free, aid for practice.

\section{Conclusions}
\label{sec:conclude}

Introducing topological data analysis, and specifically the BallMapper algorithm of \cite{dlotko2019ball}, to simple datasets on financial defaults offers much to understanding the contribution of individual financial ratios to liquidation and bankruptcy. Mapping the financial ratio characteristic space it is shown that failure only occurs within a subset of the space. It would therefore be beneficial to go further to identify that space using interaction terms. This paper has shown that interactions between liquidity, profitability, productivity, leverage and asset turnover should be explored further. BM has the ability to signpost exactly where firms are in the space, how close they are to areas where failure has been observed, and how we might then understand the decision to give credit based thereupon. Placing firms on the ``map'' is an important step to evaluating creditworthiness. Contributions of a new approach and demonstrations of non-linearity are clear. Because everything that appears in the plot is driven by the data there is no ``black-box'' criticism to any of the results that emerge; this sets BM apart from the machine learning literature which has yet to be fully trusted in practice.

A number of potential extensions emerge, with applications to other datasets, addition of further axes and consideration of wider time-frames being obvious next steps. However, each of these is just a small increment relative to the demonstration of the power of the method. As done by other works in the field these can all be left to future studies to explore. BM graphs are a map of the data cloud and hence have potential to guide the segmentation thereof, this may be a fruitful line of enquiry for subsequent research if it can be understood that there are no chances of failures occurring in any disregarded part of the space. At this stage shrinkage of the data cloud on the back of not having observed any bankruptcy outcomes would be premature. Notwithstanding TDA BM represents a new system that offers a great amount to the study of credit. This paper takes a critical first step on that representational journey. 
 
\bibliography{altman}
\bibliographystyle{apalike} 
\end{document}